\documentclass[aps,prl,twocolumn,amsmath]{revtex4-1}

\usepackage{graphicx,amsmath,amsbsy,amsthm,amsfonts,amssymb}

\newcommand{\rb}{\mathbf{r}}
\newcommand{\pb}{\mathbf{p}}

\newcommand{\Ob}{\mathbf{O}}
\newcommand{\Pb}{\mathbf{P}}

\begin{document}

\title{Quantitative imaging of anisotropic material properties with vectorial ptychography}
\author{Patrick Ferrand, Arthur Baroni, Marc Allain, Virginie Chamard}
\address{Aix-Marseille Univ, CNRS, Centrale Marseille, Institut Fresnel, F-13013 Marseille, France}
\date{\today}

\begin{abstract}
Following the recent establishment of the formalism of vectorial ptychography [Ferrand \emph{et al.}, Opt. Lett. 40, 5144 (2015)], first measurements are reported in the optical range, demonstrating the capability of the proposed method to map the four parameters of the Jones matrix of an anisotropic specimen, and therefore to quantify a wide range of optical material properties, including power transmittance, optical path difference, diattenuation, retardance, and fast-axis orientation.
\end{abstract}

\maketitle


Ptychography is a powerful imaging technique aiming at reconstructing numerically the transmission properties of an object, in both amplitude and phase \cite{Faulkner04}. In optical microscopy, being sensitive to the phase shift induced by a specimen opens new perspectives in quantitative imaging: ptychography was used, \emph{e.~g.}, to quantify of the mass of cells \cite{Marrison13} or chromosomes \cite{Shemilt15}. However, since the probe-object interaction was modeled in a scalar approximation, such applications have been restricted to isotropic materials. Many systems, however, present strong birefringent properties, as in the case of biological tissues \cite{Wolman86}, biominerals \cite{metzler2017} or materials under stress \cite{Anthony2016SR}, so they could not be investigated by ptychography so far. In order to fill that gap, we have recently introduced a generalized vectorial formalism and derived a vectorial phase retrieval algorithm aiming at reconstructing the optical properties of an anisotropic object, from measurements taken under an appropriate set of prepared and analyzed polarization states \cite{Ferrand15}. The relevance of this approach for studying the mechanical properties of optical materials was confirmed by numerical simulations \cite{Anthony2016SPIE}. In this Letter, we report on the first experimental demonstration of vectorial ptychography, and explore the wide range of material properties that can be quantitatively imaged. A model system and a biogenic specimen are investigated.


As described previously \cite{Ferrand15}, the specimen transmission properties to be retrieved must be described by a matrix of $2\times2$ Jones maps
\begin{equation}
  \Ob (\rb) = 
  \left[
    \begin{array}{cc}
      \rho_{xx}(\rb) & \rho_{yx}(\rb) \\
      \rho_{xy}(\rb) & \rho_{yy}(\rb) 
    \end{array}
  \right],
\end{equation}
so that the vectorial exit field $\Psi_{jk}(\rb)$ is obtained by a matrix multiplication
\begin{equation}
\Psi_{jk}(\rb) = \Ob(\rb) \ \pb_{jk}(\rb),
\label{eq:matrixmultip}
\end{equation}
where $\pb_{k}$ is the $k$-th probe, e.g., the $k$-th polarization state, centered at the $j$-th scanning position $\rb_j$, and is written as a complex vector
\begin{equation}
   \pb_{jk}(\rb) = \pb_{k}(\rb-\rb_j) =
     \left[
     \begin{array}{c}
       p_{jk;x}(\rb) \\
      p_{jk;y}(\rb) 
     \end{array}
   \right].
\end{equation}
Without any \emph{a priori} knowledge of the neutral axes of the object to be imaged, our proposed general strategy is to probe the specimen by three successive linearly-polarized probes $\pb_{k}$, where $k \in \{0,45,90\}$ denotes the polarization angle in degrees in the $xy$ plane, and to analyze each produced intensity pattern along three successive linear polarization directions, denoted with the index $l \in \{0, 45, 90\}$, and defined similarly. Thus the vectorial ptychographical iterative engine (vPIE) exploits at every probe position $\rb_j$ a set of nine intensity patterns, corresponding to all the combinations of probes and analyzers.

\begin{figure}[!h]
\centering
\fbox{\includegraphics[width=.8\linewidth]{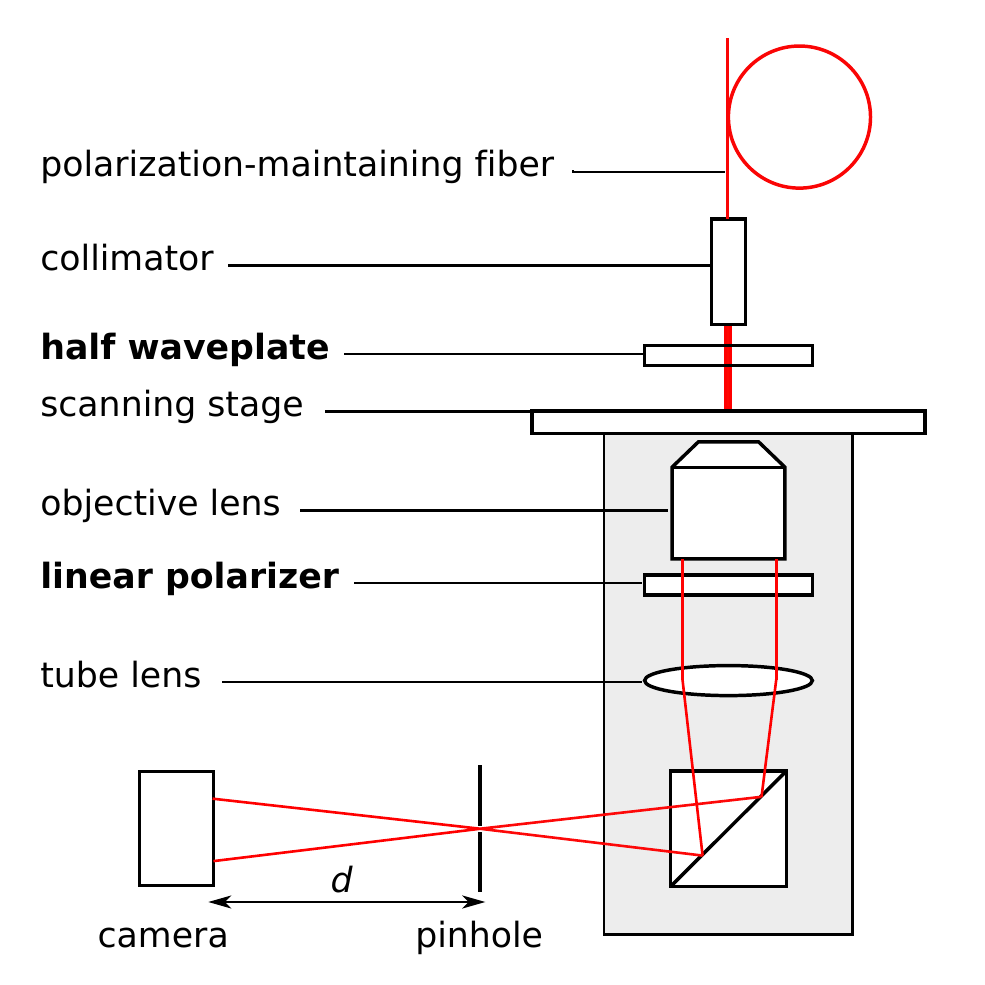}}
\caption{Scheme of the experimental bench. The specimen is placed on the scanning stage. The half waveplate and the polarizer are mounted on motorized rotation stages.}
\label{fig:Fig01}
\end{figure}

Measurements were carried out on an inverted microscope (IX73, Olympus) in a selected-area configuration \cite{maiden_SPIE_2010}, as depicted in Fig.~\ref{fig:Fig01}. The illumination was supplied by the collimated beam of a polarization-maintaining fiber-coupled laser source (S1FC635MP, Thorlabs) operating at a wavelength $\lambda = 635$~nm, mounted on the modified illumination arm. The specimen was placed on a scanning stage (U-780, Physik Instrumente). The illumination probe was optically defined by placing a 500-$\mu$m pinhole in the magnified image plane of the microscope objective lens. The diffracted intensity was recorded by a 8x8 binned CCD camera (Stingray F-145B, Allied Vision), placed at $d = 190$~mm after the pinhole.

Furthermore, the \emph{vectorial} ptychography acquisition scheme requires a full control of the polarization state of the probes, together with a proper polarization analysis. For this purpose, the linear polarization state of the illumination probe was rotated by means of a zero-order half waveplate (WPH05M-633, Thorlabs) placed after the laser collimator. Polarization analysis was performed by a linear polarizer (LPVISB100, Thorlabs) placed after the objective lens. These two polarization-control components were mounted on motorized rotation stages (K10CR1, Thorlabs) that were carefully calibrated in angle and controlled independently. Prior to every series of measurements, the background signal was measured by switching off the laser source. This signal, which includes dark noise and residual ambiant light, was found to be independent on the polarization combination, and was taken into account in the vPIE \cite{Ferrand15}.

\begin{figure}[!h]
\centering
\fbox{\includegraphics[width=\linewidth]{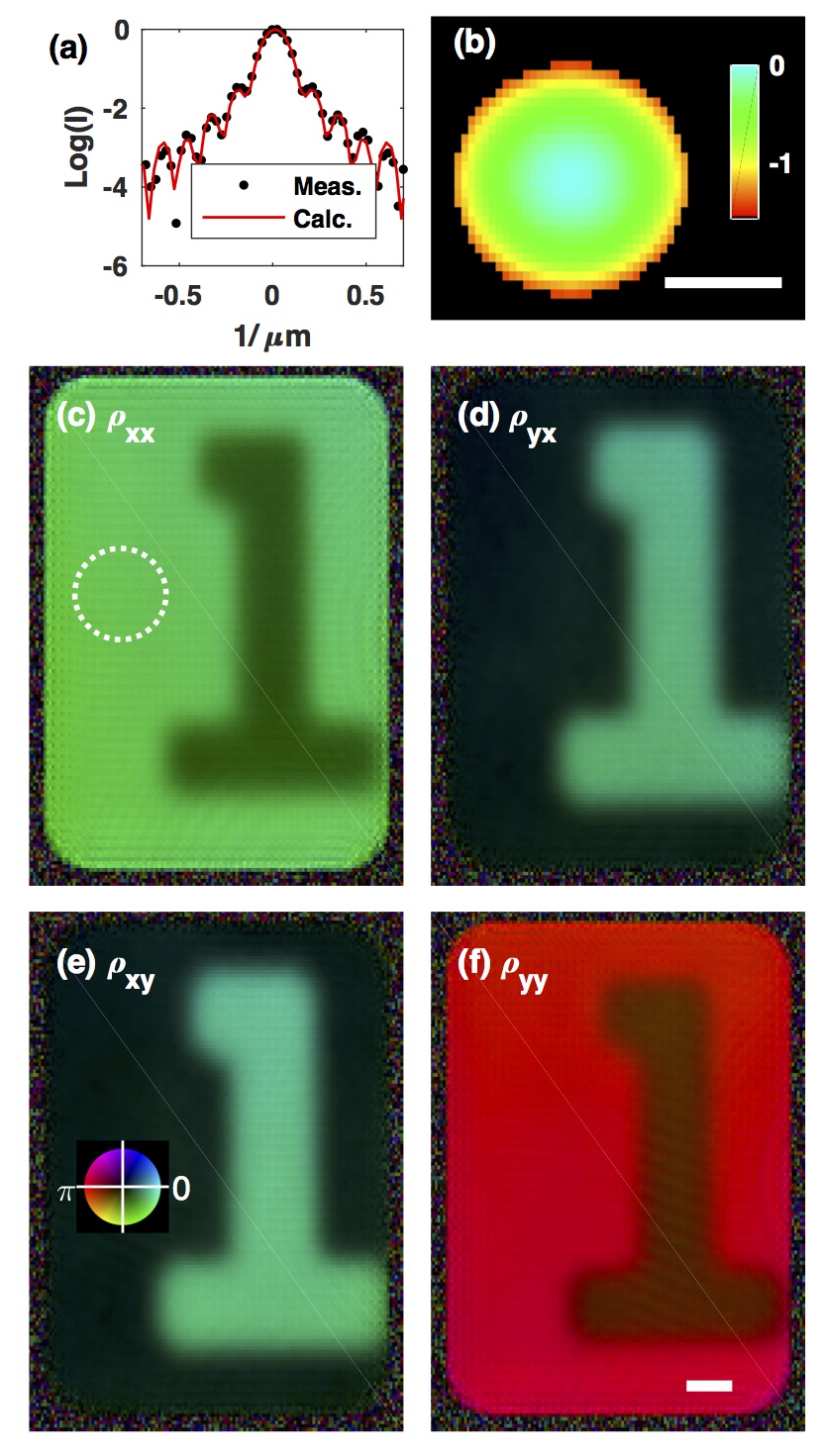}}
\caption{Reconstruction of a the NBS 1963A resolution test target. (a) Cross section of the intensity pattern recorded when illuminating a homogeneous area of the specimen, for $k=0$ and $l=0$. Dots, measurement. Solid line, fit. (b) Corresponding phase distribution (in radians) of the three probes that are used for the reconstruction. (c-f) Reconstructed Jones maps of the test target. The dotted circle represents the relative size of the probe used in the measurement. Inset shows the complex value color coding, with phase encoded as hue and modulus as brightness. All scale bars are 25~$\mu$m.}
\label{fig:Fig02}
\end{figure}

The first presented dataset were recorded on a NBS 1963A birefringent resolution target (R2L2S1B, Thorlabs). This test target was made of birefringent patterns of homogeneous retardance, and was designed so that it appeared contrasted when viewed with a dark-field polarized-light microscope, \emph{i.~e.}, when polarizer and analyzer are crossed. A 10x objective lens (ACHN-P, NA$=0.25$, Olympus) was used, making an effective demagnified probe diameter of 50~$\mu$m, under which the specimen was scanned over a 1287-point regular grid, corresponding to a raster scan with steps of 4.38 and 5.84~$\mu$m, along the $x$ and, respectively $y$ direction. Although our chosen configuration ensured a relatively good knowledge of each of our probes in amplitudes in the plane of the specimen, less was known about their phase distribution. This latter was investigated by fitting a calculated intensity pattern to an experimental one measured in a homogenous area of the specimen. The measured intensity cross section is shown in Fig.~\ref{fig:Fig02}a, for $(k,l)=(0,0)$ together with the best fit. Figure~\ref{fig:Fig02}b shows the deduced probe phase distribution, which shows a slight wavefront curvature of about 1.4~rad over the probe radius. This phase distribution was found to be the same for all three polarized probes.

The Jones maps obtained after 90 iterations of the reconstruction procedures are reported in Fig.~\ref{fig:Fig02}c. A "1" character and its homogeneous surrounding area, are clearly reconstructed. For both regions, there is unambiguous evidence of their single-layer birefringent nature, for those familiar with the Jones matrix algebra \cite{Jones41b}: (i) non diagonal terms $\rho_{yx}$ and $\rho_{xy}$ are identical, (ii) the character region exhibits a strong amplitude for these terms, which is the signature of a birefringent material having its neutral axes rotated far from the directions of the image frame; and (iii) the surrounding area shows a significant amplitude only for diagonal termes $\rho_{xx}$ and $\rho_{yy}$, but with different phases: this is the signature of a birefringent material having neutral axes oriented closely to the image frame.

In order to build a comprehensive picture of the optical properties of the object, we have solved numerically the two eigenvalues $\Lambda_{1,2}$ and two eigenvectors $\Pb_{1,2}$ of the Jones matrix $\Ob(\rb)$ at every reconstructed point $\rb$, meaning that $\Ob (\rb) \Pb_{1,2}(\rb) =\Lambda_{1,2}(\rb) \Pb_{1,2}(\rb)$. The retardance is 
\begin{equation}
R \equiv \frac{\lambda}{2\pi} \left[ (\phi_\text{slow} - \phi_\text{fast}) \pmod{2\pi} \right],
\end{equation}
where $\phi_\text{fast}$ and $\phi_\text{slow}$ refer to the phase shifts associated to the fast (respectively, slow) axis. We chose to set $(\phi_\text{fast},\phi_\text{slow}) = (\arg(\Lambda_1),\arg(\Lambda_2))$ or $(\arg(\Lambda_2),\arg(\Lambda_1))$, in order to ensure that $R$ remains in the $[0;\lambda/2]$ range. With this convention, the respective axes orientations were directly deduced from the eigenvectors components. For a first-order retardance specimen, \emph{i.~e.}, having a birefringence $\Delta n$ and a thickness $e$  so that $\Delta n e < \lambda/2$, the retrieved fast axis matches the \emph{real} fast axis of the material. For thicker specimen, it should be understood as an \emph{effective} fast axis, meaning that the materials acts optically like a first-order retarder of this fast axis orientation. Consequently, the power transmittance is $T=|\Lambda_i|^2$, the diattenuation is $\eta = \log(|\Lambda_j/\Lambda_i|)$, where $i$ and $j$ refer to the chosen fast (respectively, slow) axis-related eigenvalue. Finally, the optical phase difference is $\text{OPD} \equiv  \frac{\lambda}{2\pi} ( \phi_\text{fast} \pmod{2\pi})$.

\begin{figure}[!h]
\centering
\fbox{\includegraphics[width=\linewidth]{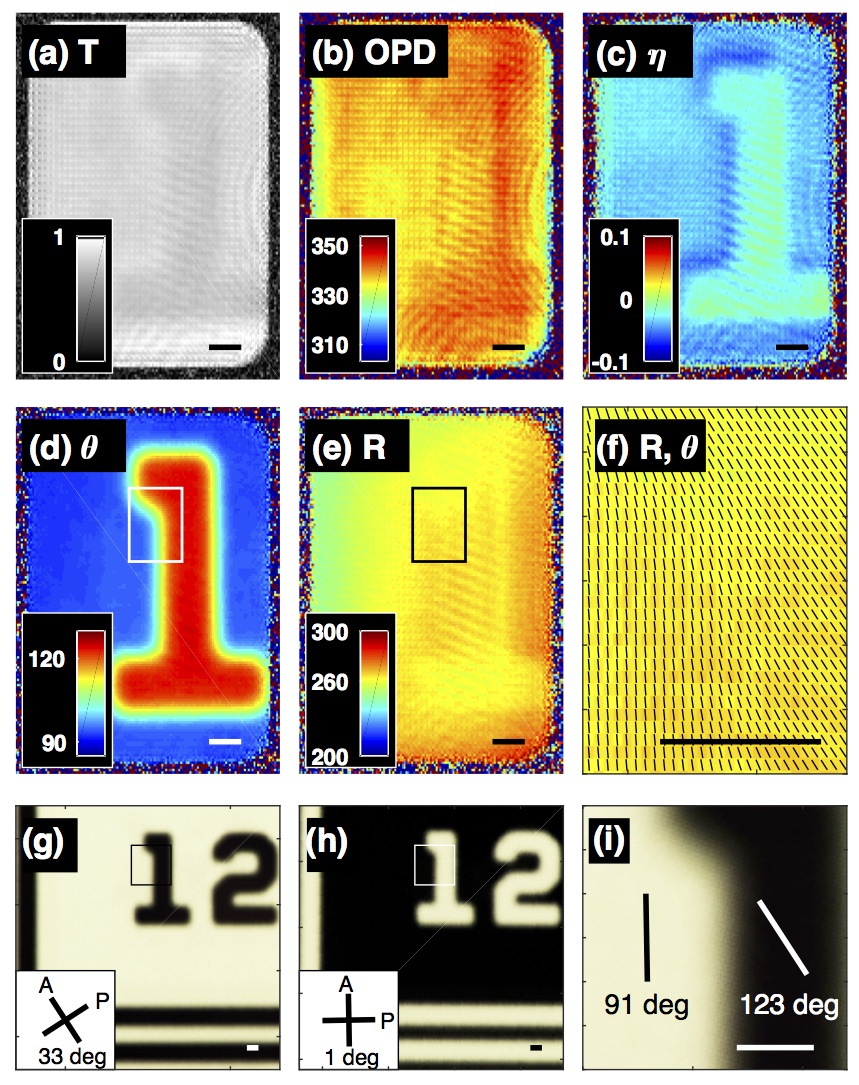}}
\caption{(a-f) Retrieved optical properties of the test target. (a) Normalized power transmittance $T$. (b) Relative OPD variations, in nm. (c) Diattenuation $\eta$. (d) Fast axis orientation $\theta$, in degrees. (e) Retardance $R$, in nm. (f) Close-up on the small rectangular region (indicated in (d) and (e)), where sticks oriented at $\theta$ are superimposed the map of retardance. (g-i) Observation of the test sample under polarized light microscopy. (g) Extinction for the characters of the target. Inset shows the corresponding orientations of the polarizer (P) and analyzer (A). (h) Extinction of the surrounding of the characters. (i) Close-up of the rectangular area of panel (e), where the obtained fast axis orientation is indicated for each region. All scale bars are 25~$\mu$m.}
\label{fig:Fig03}
\end{figure}

Results are depicted in Fig.~\ref{fig:Fig03}. Transmittance, OPD and retardance (Figs.~\ref{fig:Fig03}a, \ref{fig:Fig03}b and \ref{fig:Fig03}e, respectively) exhibit relatively flat responses. This fully agrees with the design of the target, which is made of birefringent patterns placed aside each other, and appears transparent under non polarized light. Color scales were chosen in order to emphasize very weak variations, mostly attributed to specimen imperfections and reconstruction artefacts, reproducing the scanning grid because our probes were not perfectly known. The measured retardance, in the 255--265-nm range, is in agreement with value provided by the manufacturer, $R = 280\pm20$~nm. It is interesting to notice that diattenuation (Fig.~\ref{fig:Fig03}c) underlines the overall shape of the character. The values of $\eta$ being in the $\pm0.04$ range, it corresponds to weak polarization-sensitive power transmittance change, typically in the range of $\pm 1\%$. This showed that the method is highly sensitive to the interfaces between birefringent patterns. The most striking property of the target appears clearly in Figs.~\ref{fig:Fig03}d and \ref{fig:Fig03}f. Remarkably, character and surrounding differ by their fast axis orientation, which switches from $124\pm1$~deg in the character to $92\pm1$~deg in the surrounding area. These values were confirmed by polarized light microscopy. Figures~\ref{fig:Fig03}g and \ref{fig:Fig03}g show the images obtained for the full extinction of the characters, and of the surrounding, respectively. By means of standard protocoles of polarized microscopy using a full wave plate compensator \cite{Murphy2002}, we could (i) confirm the \emph{first-order} nature of the retardance, (ii) obtain an approximate value of $R$, in the range 200--300~nm, (iii) estimate (with an uncertainty of 2~deg) the fast axis orientation in each region, as shown in Fig.~\ref{fig:Fig03}i, and (iv) confirm that the relatively low reconstructed image resolution is due to the poor definition of the target itself.

\begin{figure}[!h]
\centering
\fbox{\includegraphics[width=\linewidth]{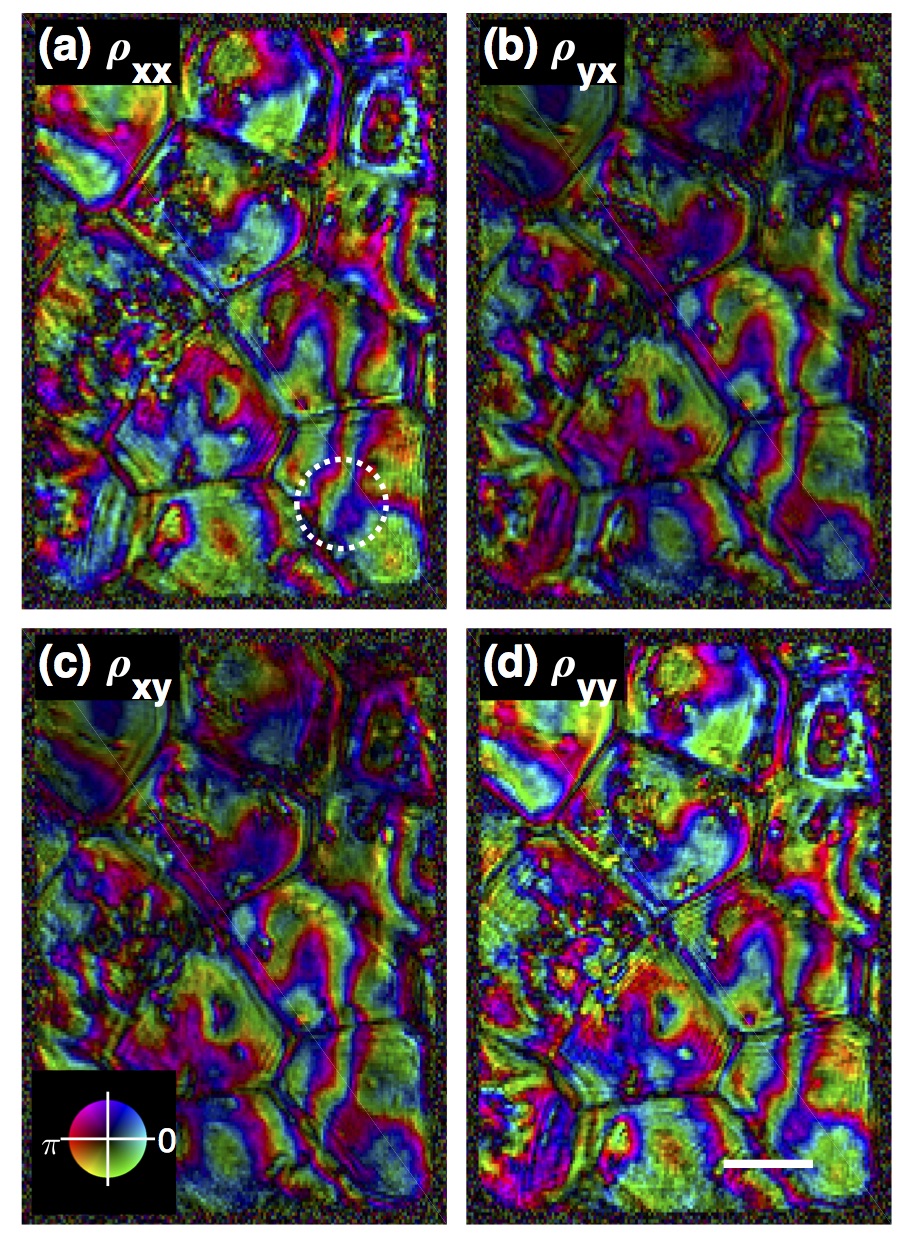}}
\caption{Reconstructed Jones map of a \emph{Pinna nobilis} oyster shell. Inset shows the complex value color coding, with phase encoded as hue and modulus as brightness. The dotted circle represents the relative size of the probe used in the measurement. Scale bar is 25~$\mu$m.}
\label{fig:Fig04}
\end{figure}

Additionally, in order to further illustrate the performance of vectorial ptychography for imaging complex systems, we have applied this method to a \emph{Pinna nobilis} oyster shell. This material constitues a famous example of biomineral, \emph{i.~e.}, a mineral produced by a living organism \cite{Cuif_biominerals_2011}, the structural properties of which are widely investigated, in order to build realistic biomineralization scenarios \cite{wolf_2013,mastropietro2017}. Our specimen was taken from the spines that cover the outer surface of the \emph{Pinna nobilis} shell \cite{Marin_2011}. Oyster shells are made of single-crystal calcite prisms separated by organic walls. Such sample is well-suited to microscopy observation because of its relatively good flatness, and moderate thickness, typically 3~$\mu$m for the present specimen. Compared with measurement reported above, we switched to a 20x objective lens (ACHN-P, NA$=0.4$, Olympus), which allow reaching a higher spatial resolution. Consequently, the demagnified probe diameter became 25~$\mu$m. The same step sizes as above were used. For this measurement, we did not notice any probe wavefront curvature, so we used a flat probe for the reconstruction. 

\begin{figure}[!h]
\centering
\fbox{\includegraphics[width=\linewidth]{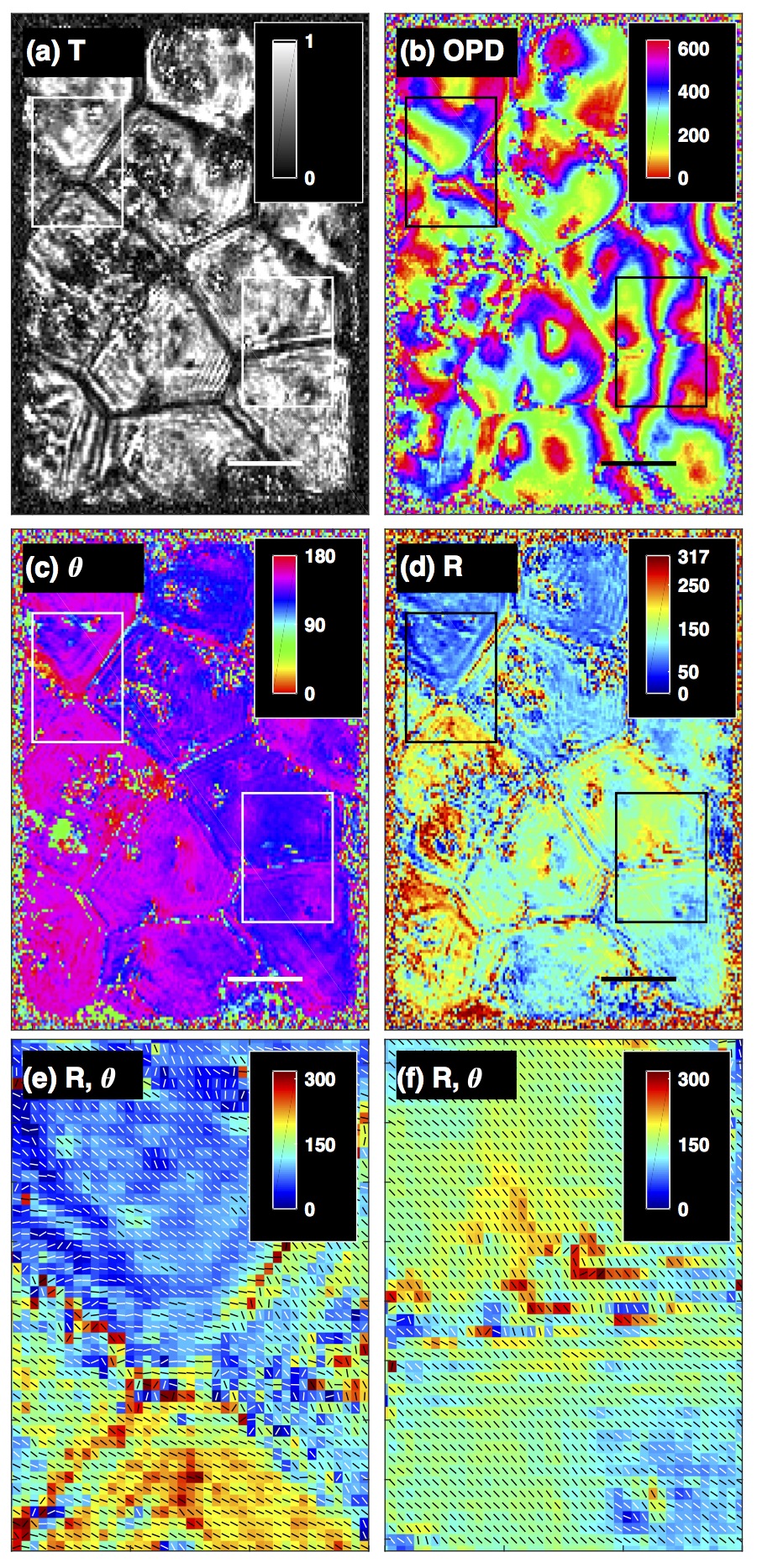}}
\caption{Retrieved optical properties of the \emph{Pinna nobilis} oyster shell. (a) Normalized power transmittance $T$. (b) Relative OPD variations, in nm. (c) Effective fast axis orientation $\theta$, in degrees. (d) Retardance $R$, in nm. (e,f) Close-ups on the left (respectively, right) rectangle area (indicated in pannels (a-d)), where sticks oriented at $\theta$ are superimposed the map of retardance. All scale bars are 25~$\mu$m.}
\label{fig:Fig05}
\end{figure}

Jones maps obtained after 90 reconstruction steps are shown in Fig.~\ref{fig:Fig04}. The micro structure of the shell, made of prisms, delimited by darker walls, appeared clearly. Remarkably, non diagonal maps $\rho_{xy}$ and $\rho_{yx}$ were identical, which was, here again, evidence for the single-layer birefringent nature of this specimen. These observations were well complemented by the computation of the optical parameters, which are summarized in Fig.~\ref{fig:Fig05}. Transmittance map (Fig.~\ref{fig:Fig05}a) corresponded to observations under conventional non-polarized microscopy. Much more quantitative information was provided by the map of OPD variations (Fig.~\ref{fig:Fig05}b), which showed smooth fluctuations of the OPD within a prism, in an approximate range of one wavelength. One can clearly observe evidence for different types of OPD behavior between adjacent prismes. One area (left rectangle) showed jumps of OPD, suggesting different structures for the two prisms, while another region (right rectangle) showed a clear OPD continuity, suggesting that two crystalline prisms shared the same optical structure, although separated by an organic wall. These observations were confirmed by the detailed analysis of the effective-fast-axis-orientation (Fig.~\ref{fig:Fig05}c) and of the retardance (Fig.~\ref{fig:Fig05}d). The closeups on the two rectangular areas (Figs.~\ref{fig:Fig05}e-f) show the high level of details that are present in this complex material system at the microscopic scale and that have been revealed by our vectorial approach. In comparison, polarized light microscopy would have required to distinguish between more than fifty shades of grey.

In conclusion, we have reported the first experimental demonstration of vectorial ptychography. Measurements performed on a test sample allowed to confirm the power of the method in quantifying the anisotropic optical parameters. Measurements performed on a biomineral specimen confirmed its ability to address complex specimens.

\section*{Fundings}

This work has received fundings from Institut Fresnel, from the French Agence Nationale de la Recherche under contract no ANR-11-BS10-0005, and from the European Research Council (ERC) under the European Union’s Horizon H2020 research and innovation program grant agreement No 724881.

\section*{Acknowledgments}

The specimen of \emph{Pinna nobilis} was provided by courtesy of Prof. J.-P. Cuif, who is gratefully acknowledged.

\bibliography{ptycho}

\end{document}